\documentstyle[12pt,aps,prd,preprint,graphicx]{revtex}
\begin{document}
\title{High-Order Contamination in the Tail of Gravitational Collapse}
\author{Shahar Hod}
\address{The Racah Institute for Physics, The
Hebrew University, Jerusalem 91904, Israel}
\date{\today}
\maketitle

\begin{abstract}

It is well known that the late-time behaviour of gravitational
collapse is {\it dominated} by an inverse power-law decaying tail. 
We calculate {\it higher-order corrections} to this power-law behaviour in a 
spherically symmetric gravitational collapse. 
The dominant ``contamination'' is shown to die off
at late times as $M^2t^{-4}\ln(t/M)$. This decay rate is 
much {\it slower} than has been considered so far. It implies, for instance,
that an `exact' (numerical) determination of the power index to within
$\sim 1 \%$ requires extremely long integration times of order $10^4 M$. 
We show that the leading order fingerprint of the black-hole 
electric {\it charge} is of order $Q^2t^{-4}$.
\end{abstract}

\section{introduction}\label{introduction}

Waves propagating in a curved spacetime develop ``tails''. 
In particular, it is well established that the {\it dominant} late-time behaviour of
massless fields propagating in black-hole spacetimes is a power-law
tail. 
Price \cite{Price} was the first to analyze the mechanism by which the
spacetime outside a (nearly spherical) collapsing star divests itself of
all radiative multipole moments, and leaves behind a Schwarzschild
black hole; it was demonstrated that all radiative perturbations decay
asymptotically as an inverse power of time. 
Physically, these inverse
power-law tails are associated with the backscattering of waves off
the effective curvature potential at asymptotically far regions
\cite{Thorne,Price}.

The analysis of Price was extended by other authors. Bi\v{c}\'{a}k
\cite{Bicak} generalized the analysis and studied the dynamics of a 
scalar field in a {\it charged} Reissner-Nordstr\"om spacetime. He also
found an asymptotic inverse power-law decay of the field, with the {\it same}
power indices as in the Schwarzschild spacetime (with the exception of
the {\it extremal} Reissner-Nordstr\"om black hole \cite{Bicak}). In a
brilliant work, Leaver \cite{Leaver} demonstrated that the late-time tail can be
associated with the existence of a branch cut in the Green's function
for the wave propagation problem. 

Gundlach, Price, and Pullin \cite{Gundlach1} showed that these inverse
power-law tails also characterize the late-time evolution of radiative
fields at future null infinity, and at the black-hole outer
horizon. Furthermore, they showed that 
power-law tails are a genuine feature of gravitational collapse --
the existence of these tails was demonstrated in full non-linear 
numerical simulations of the spherically symmetric collapse 
of a self-gravitating scalar 
field \cite{Gundlach2} (this was later reproduced in \cite{BurOr}).

Our current understanding of the late-time tail is, however, somewhat
unsatisfactory. The ({\it leading order}) power-law tails in
black-hole spacetimes are well established 
\cite{Price,Thorne,Bicak,Leaver,Gundlach1,Gundlach2,BurOr,Ching,Andersson1,Barack},
but the resultant formulae
are only truly useful at very {\it late} times. In a typical evolution
scenario there is a considerable time window in which the signal is no
longer dominated by the quasinormal modes 
\cite{Leaver,SuPr,Gundlach1,Gundlach2,Andersson2,Ching,Andersson1}, 
but the leading order power-law tail has not yet taken over \cite{Andersson1}.

The purpose of this paper is to derive
analytic expressions for the {\it higher-order corrections} which
``contaminate'' the well-known power-law tail in a spherically symmetric 
gravitational collapse. The determination of
these higher-order terms is important from several points of view: 
The analyses of Bi\v{c}\'{a}k \cite{Bicak} and 
Gundlach et. al. \cite{Gundlach1} established the fact that the
leading-order power-law tail is {\it universal} in the sense that it
is {\it independent} of the black-hole
electric charge (i.e., the power index in a {\it charged}
Reissner-Nordstr\"om spacetime was shown to be identical with 
the one found by Price \cite{Price} for the neutral Schwarzschild
black hole). 
This observation begs the question: what fingerprints (if any) does
the black-hole electric charge leave on the field's decay ?

Moreover, the calculation of higher-order corrections to the leading
order power-law tail is also of practical importance; this is
  especially crucial for the determination of the power index from
  numerical simulations. The dominant inverse power-law tail is
  {\it ``contaminated''} by higher-order terms, whose effect become larger
  as the aveliable time of integration decreases. The precise power
  index is expected only at infinitely-late time. Thus, in practice, 
the {\it limited} time of integration introduces an inherent error in
the determination of the power index. The only systematic approach to
{\it quantify} the errors which are introduced by the finite integration time
is to study {\it higher-order corrections}. If one computes the
contaminated part of the late-time tail, then the ratio of the
corrections to the leading order term is a systematic, quantitative,
indication of the error caused by the {\it finite}-time numerical calculation.

These questions and several others are addressed in the present paper. 
The plan of the paper is as follows. In Sec. \ref{Sec2} we give a
short description of the physical system and formulate the evolution
equation considered. In Sec. \ref{Sec3} we give an analytical
description of the late-time evolution of scalar fields 
in black-hole spacetimes. In Sec. \ref{Sec4} we confirm our analytical results
by numerical simulations. 
We conclude in Sec. \ref{Sec5} with a brief summary of our results and
their implications.

\section{Description of the system}\label{Sec2}

We consider the evolution of a spherically symmetric massless scalar field in a
spherically symmetric charged background (a collapsing star or a fixed
black hole). 
The external gravitational field of a spherically symmetric charged object of
mass $M$ and charge $Q$ is given by the Reissner-Nordstr\"om metric

\begin{equation}\label{Eq1}
ds^2=-\left( {1-{{2M} \over r}+{{Q^2} \over {r^2}}} \right)dt^2+\left( {1-{{2M}
\over r}+{{Q^2} \over {r^2}}} \right)^{-1}dr^2+r^2d\Omega ^2\  .
\end{equation} 
Using the tortoise radial coordinate $y$, which is 
defined by $dy=dr/(1-2M/r+Q^2/r^2)$, the line element becomes

\begin{equation}\label{Eq2}
ds^2=\left( {1-{{2M} \over r}+{{Q^2} \over {r^2}}} \right)\left( {-dt^2+dy^2}
\right)+r^2d\Omega ^2\  ,
\end{equation}
where $r=r(y)$.

The wave equation $\phi_{;ab}g^{ab} \phi =0$ for the scalar field in
the black-hole background is 

\begin{equation}\label{Eq3}
\psi _{,tt}-\psi _{,yy}+V\psi =0\  ,
\end{equation}
where

\begin{equation}\label{Eq4}
V=V_{M,Q}(r)=\left( {1-{{2M} \over r}+{{Q^2} \over {r^2}}}
\right)\left( {{2M} \over {r^3}}-{{2Q^2}
\over {r^4}} \right)\  .
\end{equation}
In terms of the  tortoise coordinate $y$ and for $y \gg M$ 
the curvature potential Eq. (\ref{Eq4}) reads

\begin{equation}\label{Eq5}
V=V(y)={{2M} \over {y^3}}+ {{12M^2\ln y} \over {y^4}} -{{4M^2+2Q^2}
  \over {y^4}} +O \Bigg({{M^3\ln y} \over {y^5}} \Bigg)\  .
\end{equation}

\section{The late-time behaviour}\label{Sec3}

The general solution to the wave-equation (\ref{Eq3}) can be written
as a series depending on two arbitrary functions $F$ and $G$ \cite{Price}

\begin{eqnarray}\label{Eq6}
\psi& = & {{G^{(0)}(u)+F^{(0)}(v)}}\nonumber \\
 &&+  \sum\limits_{k=0}^\infty  {B_k(y)
\left[ G^{(-k-1)}(u)+({-1})^kF^{(-k-1)}(v) \right]}\  .
\end{eqnarray}
Here $u\equiv t - y$ is a retarded time coordinate and $v \equiv t + y$ is an advanced
time coordinate. For any function $H$, $H^{(k)}$ is the $k$th
derivative of $H^{(0)}$; 
negative-order derivatives are to be interpreted as
integrals. The first two terms in Eq. (\ref{Eq6}) represent the
zeroth-order solution (with $V=0$).

The star begins to collapse at a retarded time $u=u_0$. The world line of the
stellar surface is asymptotic to an ingoing null line $v=v_0$, while the
variation of the field on the stellar surface is asymptotically infinitely
redshifted \cite{Price,Bicak}. This effect is caused by the time dilation
between static frames and infalling frames. A static external observer sees all
processes on the stellar surface become ``frozen'' as the star approaches the
horizon. Thus, he sees all physical quantities approach a constant. We
therefore make the explicit assumption that after some retarded time
$u=u_1, \partial _u\phi=0$ on $v=v_0$. This assumption has been proven
to be very successful \cite{Price,Gundlach1,HodPir1}.

We begin with the first stage of the evolution, i.e., the scattering
of the field in the region $u_0 \leq u \leq u_1$. 
The first two terms in Eq. (\ref{Eq6}) represent the primary waves in the wave front,
while the sum represents backscattered waves. The interpretation of
these integral terms as backscatter comes from the fact that they depend on
data spread out over a {\it section} of the past light cone, while outgoing
waves depend only on data at a fixed $u$ \cite{Price}.

After the passage of the primary waves there is no outgoing radiation for $u >
u_1$, aside from backscattered waves. This means that $G(u_1) = 0$. Hence, for a
large $y$ at $u = u_1$, 
the dominant term in Eq. (\ref{Eq6}) is $\psi(u=u_1,y)=B_0(y)G^{(-1)}(u_1)$.

The functions $B_k(r)$ satisfy the recursion relation
$2B^{'}_{k}-B^{''}_{k-1}+VB_{k-1} =0$ for $k>1$, 
where $B'\equiv dB/dy$, and $B^{'}_{0}=-V(y)/2$. Thus, one finds

\begin{equation}\label{Eq7}
\psi (u=u_1,y)=\left[My^{-2}/2+2M^2y^{-3}\ln y-Q^2y^{-3}/3
\right]G^{(-1)}(u_1) \Big[1+O(M/y) \Big]\  .
\end{equation}
This is the dominant backscatter of the primary waves.

With this specification of characteristic data on $u=u_1$, we shall
next consider the asymptotic evolution of the field. We confine our
attention to the region $u>u_1$, $y \gg M, |Q|$. In this region the
spacetime is approximated as flat \cite{Price,Gundlach1}. (The validity of this approximation is 
ultimately justified by numerical simulations). 
Thus, the solution for $\psi$ can be written as

\begin{equation}\label{Eq8}
\psi=  {{g^{(0)}(u)+f^{(0)}(v)}}\  .
\end{equation}
Comparing Eq. (\ref{Eq8}) with the initial data on $u=u_1$ Eq. (\ref{Eq7}), one
finds

\begin{equation}\label{Eq9}
f(v)=F_0v^{-2} +F_1v^{-3}\ln v+F_2v^{-3}\  ,
\end{equation}
where

\begin{equation}\label{Eq10}
F_0=2MG^{(-1)}(u_1) \ , \quad F_1=16M^2G^{(-1)}(u_1) \ , \quad
F_2=-8Q^2G^{(-1)}(u_1)/3\  .
\end{equation}

For late times $t \gg y$ we can expand
$g(u)=\sum\limits_{n=0}^{\infty} (-1)^ng^{(n)}(t)y^n/n!$ and similarly
for $f(v)$. Using these expansions we can rewrite Eq. (\ref{Eq8}) as

\begin{equation}\label{Eq11}
\psi =\sum\limits_{n=0}^{\infty}  {K_0^ny^n\left[ f^{(n)}(t)+(-1)^ng^{(n)}(t) \right]}\  ,
\end{equation}
where the coefficients $K_0^n$ are those given in \cite{Price}.

Using the boundary conditions for small $r$ (regularity as $y \to -\infty$, at
the horizon of a black hole, or at $r=0$ for a stellar model), one finds that
at late times the terms $h(t)\equiv f(t)+g(t)$ and $f^{(1)}(t)$ must be of the
same order (see \cite{Gundlach1} for additional details). Thus, we conclude that

\begin{equation}\label{Eq12}
f(t) \simeq F_0t^{-2} +F_1t^{-3}\ln t+F_2t^{-3} \ , \quad 
g(t) \simeq -\left(F_0t^{-2} +F_1t^{-3}\ln t \right)\  ,
\end{equation}
and

\begin{equation}\label{Eq13}
h(t) \equiv f(t)+g(t)=O(M^2t^{-3})\  .
\end{equation}
We therefore find that the late-time behaviour of the field for 
$t \gg y \gg M,|Q|$ is

\begin{eqnarray}\label{Eq14}
\psi &\simeq& 2K_0^1yf^{(1)}(t) \nonumber \\
&=&-2K_0^1y\left[2F_0t^{-3}+3F_1t^{-4}\ln(t/M)+3F_2t^{-4}
\right] \Big[1+O(M/t)\Big] \  . 
\end{eqnarray}
This is the late-time behaviour of the field at a fixed radius.

\section{Numerical results}\label{Sec4}

It is straightforward to integrate Eq. (\ref{Eq3}) using the method
described in \cite{Gundlach1}. We have used, however, a modified
version of the numerical code used in \cite{HodPir2}, which is
essential to achieve the extremely high accuracy needed for the
computation (see \cite{HodPir2} for additional details).

The late-time evolution of the scalar
field is independent of the form of the initial data used. The
results presented here are for a Gaussian pulse on $u=0$
\begin{equation}\label{Eq15}
\psi (u=0,v)= A \exp \left \{-\left [ (v-v_{0})/ \sigma \right ]^{2}
\right \}\  ,
\end{equation}
with a center at $v_0=10$ and a width
$\sigma=2$. The black-hole mass is set equal to $M=0.5$; 
this corresponds to the freedom to rescale the coordinates by an
overall length scale.

\begin{figure}
\centering
\noindent
\includegraphics[width=17cm]{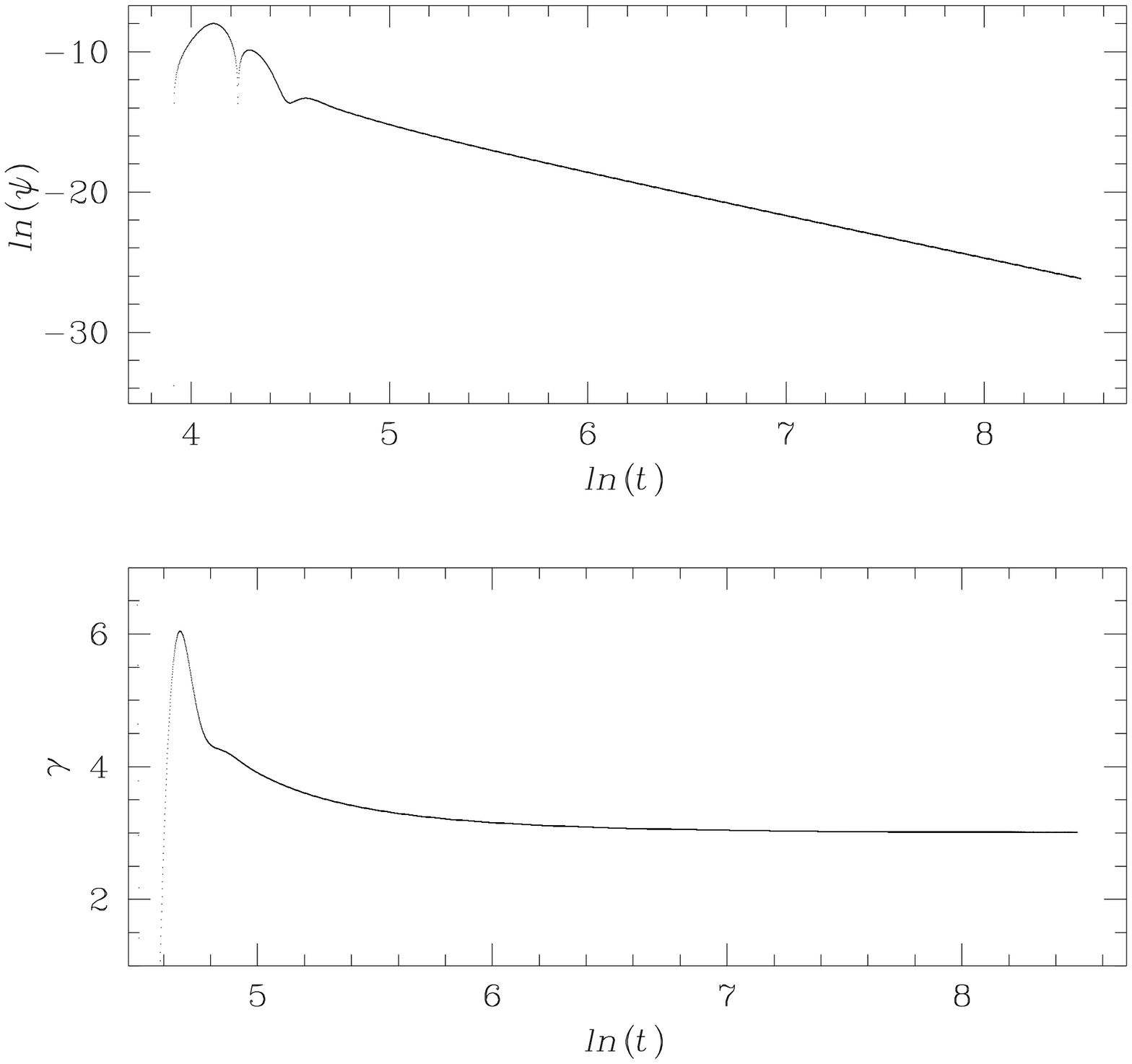}
\caption[gamma]{\label{Fig1}
Temporal evolution of the scalar field, evaluated at $y=50$ in a
Schwarzschild spacetime with $M=0.5$. The initial data is
a Gaussian distribution with $v_{0}=10$ and $\sigma =2$.  
The {\it asymptotic} power-law fall off is manifest at late times (top
panel). The bottom panel depicts the evolution of the local 
power index $\gamma \equiv -t\psi_t/\psi$. The power index
approaches the well-known asymptotic 
value $\gamma_{asy}=3$ at late times.}
\end{figure}

The temporal evolution of the field at a fixed radius
$y=50$ is shown in the top panel of Fig. \ref{Fig1}. 
The dominant {\it power-law} fall off is manifest at
asymptotic late times. 
In order to study the contamination effect of higher-order terms
[see Eq. (\ref{Eq14})], we use the notion 
of a {\it local power index} $\gamma$, defined by $\gamma \equiv
-t\psi_t/\psi$ \cite{BurOr}. Taking cognizance of Eq. (\ref{Eq14}) we find

\begin{equation}\label{Eq16}
\gamma=3+12{M \over t} \ln (t/M) +O(M/t)\  .
\end{equation}
The approach of the local power index to 
its well-known asymptotic value $\gamma_{asy} \to 3$ is depicts in the bottom panel
of Fig. \ref{Fig1}. The plot shows that $\gamma \to 3$ from above,
with a qualitative agreement with Eq. (\ref{Eq16}).

\begin{figure}
\centering
\noindent
\includegraphics[width=17cm]{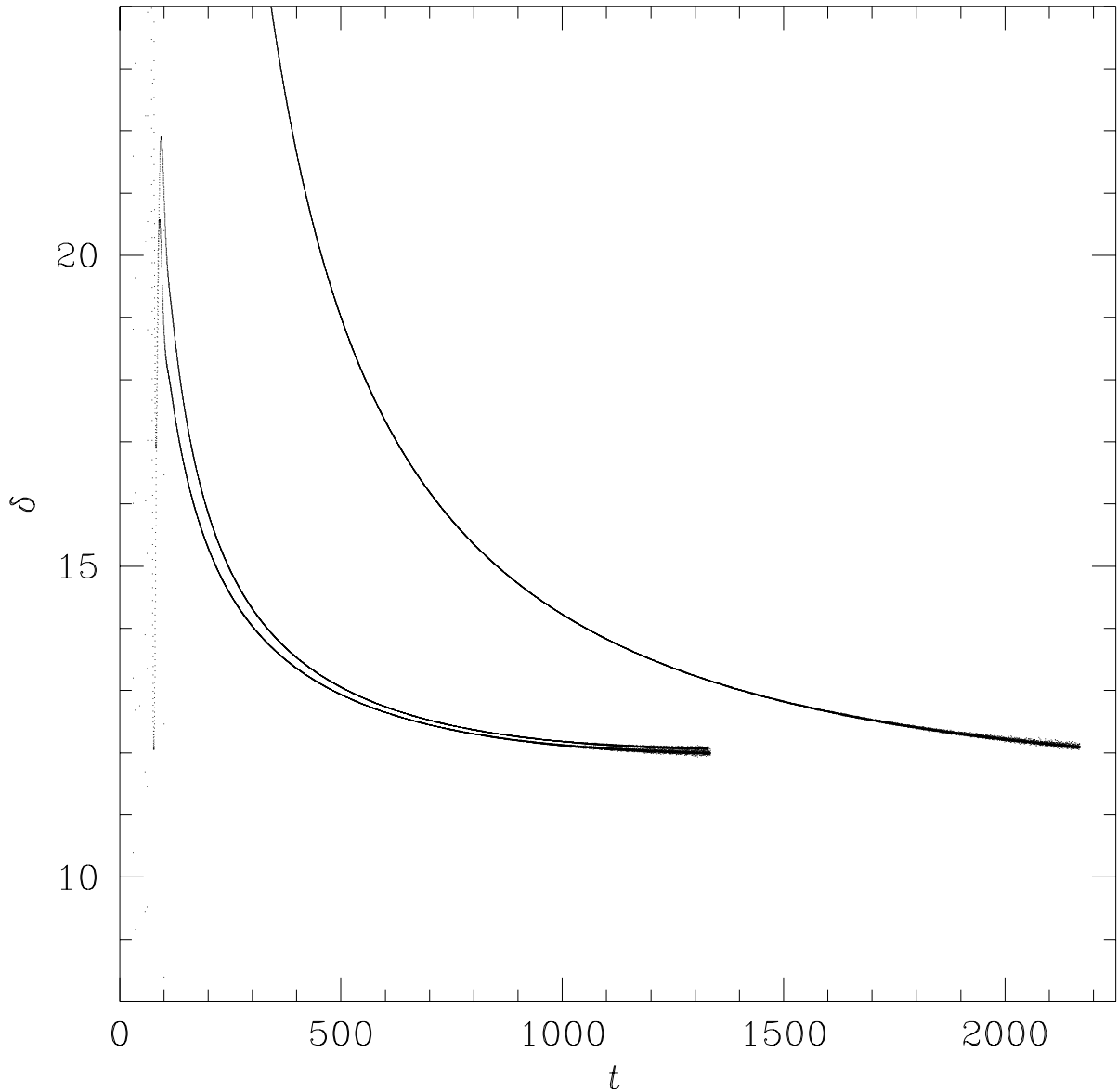}
\caption[delta]{\label{Fig2}
The time evolution of the quantity $\delta \equiv (\gamma
-\gamma_{asy})(t/M)/\ln(t/M)$, evaluated at $y=5,10$, and $50$ 
(from bottom to top). The asymptotic value 
$\delta_{asy} \simeq 12$ is in excellent agreement with 
the {\it analytically} predicted behaviour $\delta \to 12$ at late times. This
result establishes the existence of the contamination 
term of order $M^2t^{-4}\ln(t/M)$.}
\end{figure}

In order to establish {\it quantitatively} the physical picture
presented in Sec. \ref{Sec3}, we define the quantity $\delta \equiv (\gamma
-\gamma_{asy})(t/M)/\ln(t/M)$. Figure \ref{Fig2}. depicts 
$\delta$ as a function of $t$ at three surfaces of constant radius
$y=5, 10$, and $50$ (from bottom to top). The numerical 
result $\delta_{asy} \simeq 12$ (independently of the value of $y$) 
is in excellent agreement with the {\it analytically} predicted
behaviour $\delta \to 12$ [see Eq. (\ref{Eq16})]. Thus,
Fig. \ref{Fig2}. establishes the existence of the contamination 
term of order $M^2t^{-4}\ln(t/M)$. It should be noted that the value of
$\gamma_{asy}$ used in this figure is $2.998$ rather than the
theoretical value $3$. This slight deviation
from the theoretical value is expected due to the corrections of order
$O(M/t)$ in the expression for $\gamma$ Eq. (\ref{Eq16}). 

\section{Summary and physical implications}\label{Sec5}

The purpose of this paper was to derive
analytic expressions for the {\it higher-order corrections} which
``contaminate'' the well-known power-law tail in a spherically symmetric 
gravitational collapse. We have shown, both analytically and
numerically, that the dominant correction dies off
at late times as $M^2t^{-4}\ln(t/M)$. This late-time decay of the
contamination is much {\it slower} than has been considered so far 
\cite{Andersson1} (see the discussion in Appendix B). 

Aside from being {\it theoretically} important, the result
Eq. (\ref{Eq14}) is also of {\it practical} importance. 
It follows that an `exact' (numerical) determination of the
power index demands extremely long integration times. The most
accurate method for determining the power index experimentally applies
to the concept of the {\it local} power index. In this way one
discard the relatively large contamination which characterizes the
early stages of the evolution. Still, it follows from Eq. (\ref{Eq16})
that a determination of the power index to within $\sim 1 \%$ requires an
integration time of order $t \gtrsim 3000 M$.

The dominant power-law tail is known to be universal in the sense that
it is {\it independent} of the black-hole parameters
\cite{Bicak,Gundlach1}. We have shown, however, that this universality
is removed once we consider higher-order corrections terms -- 
the leading order fingerprint of the black-hole 
electric {\it charge} behaves as $Q^2t^{-4}$.

\bigskip
\noindent
{\bf ACKNOWLEDGMENTS}
\bigskip

I thank Tsvi Piran for discussions. 
This research was supported by a grant from the Israel Science Foundation.

\appendix

\section{Late-time behaviour at null infinity and at the horizon}

It follows from Eqs. (\ref{Eq8}), (\ref{Eq9}), and (\ref{Eq12}) that
the asymptotic behaviour of the field at future null infinity $scri_+$ (i.e., at
$v \gg u$) is

\begin{equation}\label{EqA1}
\psi (v \gg u,u) \simeq g^{(0)}(u) \simeq -\left(F_0u^{-2}
  +F_1u^{-3}\ln u \right)\  .
\end{equation}

We finally consider the behaviour of the field at the
black-hole outer horizon $r_+$. As $y \to - \infty$ the curvature
potential Eq. (\ref{Eq4}) is exponentially small, and the general
solution to Eq. (\ref{Eq3}) can be written as
$\psi=\alpha(u)+\gamma(v)$. 
On $v = v_0$ we take $\partial_u\phi=0$ (for $u \to \infty$). Thus,
$\alpha(u)$
must be a constant, and with no loss of generality we can choose it to
be zero. We next expand $\gamma(v)$ for $t \gg |y|$ as

\begin{equation}\label{EqA2}
\psi =\gamma (v)=\sum\limits_{n=0}^\infty  {{1 \over
{n!}}\gamma ^{(n)}(t)y^n}\  .
\end{equation}
In order to match the $y \ll -M$ solution Eq. (\ref{EqA2}) with the $y
\gg M$ solution Eq. (\ref{Eq14}), we make the ansatz $\psi \simeq
\psi_{stat}(y)
\left[2F_0t^{-3}+3F_1t^{-4}\ln(t/M)+3F_2t^{-4}\right]$ for the
solution
in the region $y \ll -M$ and $t \gg |y|$. In other words, we assume
that the
solution in the $y \ll -M$ region has the same late time $t$
dependence as the
$y \gg M$ solution. This assumption has
been proven to be very successful for the leading order behaviour of
both neutral \cite{Gundlach1} and charged \cite{HodPir1} fields. 
Using this assumption, one finds 
$\gamma(t)=\Gamma_0
\left[2F_0t^{-3}+3F_1t^{-4}\ln(t/M)+3F_2t^{-4}\right]$, 
where $\Gamma_0$ is a constant. 
Thus, the asymptotic behaviour of the field at the black-hole 
horizon is 

\begin{equation}\label{EqA3}
\psi (u \to \infty ,v)=\Gamma_0
\left[2F_0v^{-3}+3F_1v^{-4}\ln(v/M)+3F_2v^{-4}\right]\  .
\end{equation}

\section{A note on Andersson's analysis}

The first attempt to calculate higher-order corrections to the
dominant power-law tail was made by 
Andersson \cite{Andersson1} (in the context of the Schwarzschild
spacetime). This analysis was based on an {\it approximated} curvature
potential in the region far away from the black hole 
[see Eq. (26) of \cite{Andersson1}]. While this approximated
potential simplifies the analysis, it actually {\it misses} the
genuine leading order corrections terms. The leading order correction in
\cite{Andersson1} was found to be of order $Mt^{-5}$. This (artificial)
result is caused by the fact that terms of order $M/r^3$ 
(and smaller) were totally neglected by the approximated curvature potential 
used in \cite{Andersson1}.

Moreover, the approximated
approach presented in \cite{Andersson1}, {\it if} extended along the
same lines to a Reissner-Nordstr\"om spacetime would imply that
the influence of the black-hole electric charge on the late-time tail
vanishes {\it identically} (i.e., it would vanish to {\it any} order
in $t^{-1}$). This result is again a direct consequence of the fact
that the approximated approach of \cite{Andersson1} does not take
into account curvature terms of order $Q^2/r^4$ (and $M^2/r^4$) 
which appear in the (exact) curvature potential.

\end{document}